\documentclass{article}

\usepackage{graphicx}
\usepackage{bm}
\usepackage{amsmath}
\usepackage{cuted}
\usepackage{verbatim}
\usepackage{arxiv}
\newcommand{\norm}{c}			
\newcommand{\weight}{W}			
\newcommand{\Emean}{\mu}   		
\newcommand{\vv}[2][]{{\mathbf{#2}_{#1}}}
\newcommand{\vc}[2][]{{\overline{\vv{#2}}_{#1}}}
\newcommand{\vd}[2][]{{\overline{\vv{#2}}_{#1}}^T}
\newcommand{\DD}{\vv{\Delta}}
\newcommand{\Dc}{\vc{\Delta}}

\renewcommand{\aa}[1][]{\vv[#1]{a}}
\newcommand{\ac}{\vc{a}}
\newcommand{\ad}{\vd{a}}
\newcommand{\af}{\vv{a}_*}
\newcommand{\afc}{\vc{a}_*}
\newcommand{\afd}{\vd{a}_*}
\newcommand{\ww}{\vv{w}}
\newcommand{\wc}{\vc{w}}
\newcommand{\wwd}{\vd{w}}
\newcommand{\wf}{\vv{w}_*}

\newcommand{\wfd}{\vd{w}_*}
\newcommand{\ff}{\vv{f}}

\newcommand{\TM}{\vv{T}}
\newcommand{\TMd}{\vv{T}^\dag}
\newcommand{\fc}{\overline{\vv{f}}}

\newcommand{\fsuba}{{\substack{\aa=\af \\ \ac=\afc}}}

\newcommand{\pnorm}[1]{\left\lVert#1\right\rVert}
\newcommand{\valueat}[2]{\left.{#1}\right\rvert_{#2}}
\newcommand{\pderiv}[2]{\frac{\partial#1}{\partial#2}}
\newcommand{\ppderivs}[2]{\frac{\partial^2#1}{\partial#2^2}}
\newcommand{\ppderiv}[3]{\frac{\partial^2#1}{\partial#2\partial#3}}
 
\newcommand{\supfill}{\phantom{}} 

\begin{document}
\title{Blind focusing through strongly scattering media using wavefront shaping with nonlinear feedback}
\author{Gerwin Osnabrugge,\authormark{1,*} Lyubov V. Amitonova,\authormark{2} and Ivo M. Vellekoop\authormark{1}}
%
%

\author{
  Gerwin Osnabrugge \\
  Biomedical Photonic Imaging Group \\
  University of Twente \\
  7500 AE Enschede, The Netherlands \\
  \texttt{g.osnabrugge@utwente.nl} \\
   \And
  Lyubov V. Amitonova \\
  Department of Physics and Astronomy \\
  Vrije Universiteit Amsterdam \\
  1081 HV Amsterdam, The Netherlands \\
	\And
  Ivo M. Vellekoop \\
  Biomedical Photonic Imaging Group \\
  University of Twente \\
  7500 AE Enschede, The Netherlands
}
	



\maketitle

\begin{abstract}
Scattering prevents light from being focused in turbid media. The effect of scattering can be negated through wavefront shaping techniques when a localized form of feedback is available. Even in the absence of such a localized reporter, wavefront shaping can blindly form a single diffraction-limited focus when the feedback response is nonlinear. We developed and experimentally validated a model that accurately describes the statistics of this blind focusing process. We show that maximizing the nonlinear feedback signal does not always result in the formation of a focus. Using our model, we can calculate the minimal requirements to blindly focus light through strongly scattering media. 
\end{abstract}

\section{Introduction}
Refractive index inhomogeneities in a turbid medium scatter light in a complex manner. Consequently, a focus inside these types of media becomes more aberrated with increasing depth, until ultimately no ballistic light is left and the focus decays into a random speckle pattern. Even at this depth, light can still be focused through wavefront shaping techniques \cite{Vellekoop2015,Horstmeyer2015}. These techniques have been used to focus light inside scattering media for various applications, such as optical manipulation \cite{Dholakia2011}, optogenetics \cite{Ruan2017} and fluorescence microscopy through an intact skull of a mouse \cite{Park2015}. Using wavefront shaping techniques, light has been focused through several centimeters of biological tissue \cite{Shen2016}.

The main limitation of these wavefront shaping techniques is the requirement of a localized reporter (e.g. guide star), providing feedback of the light intensity at the focus location. Many different mechanisms can act as localized reporters, such as point detectors, fluorescent or nonlinear markers, acoustically tagged light and photoacoustic absorbers \cite{Horstmeyer2015}. However, a localized reporter is not always available. For instance in multiphoton fluorescence excitation microscopy, generally all structures of interest are stained with fluorescent markers, which will all generate a signal when illuminated. When the excitation light is spread over a larger area, the total feedback signal will originate from multiple indistinguishable reporters. However, when a weak ballistic unscattered component is still preserved, this mixed form of feedback can be used with wavefront shaping to correct an aberrated focus \cite{Tang2012, Sinefeld2015, Milkie2011}. 

Katz et al. \cite{Katz2014} showed that even in the absence of ballistic light, light can be `blindly' focused through a strongly scattering layer when feedback was generated by many fluorescent sources behind the opaque layer. Moreover, wavefront shaping with nonlinear feedback is not limited to continuous wave sources, but can also be employed to temporarily focus a scattered light pulse at one of the feedback sources \cite{Katz2011,Aulbach2012}. Therefore, this blind focusing technique could potentially be a powerful technique for nonlinear deep-tissue microscopy. So far, a theoretical understanding of this iterative optimization process has been missing, and, therefore, it is not known under which conditions the blind focusing method will converge to a single diffraction-limited focus.

Here, we present and validate a model that accurately describes the statistics of blind focusing. Using this model, we show that focusing is only possible when certain minimal requirements are met. Additionally, we are able to predict the evolution of the optimized speckle pattern during the optimization process.

We will first derive the exact solution of the scattered field behind a strongly scattering layer for a single (known) realization of the scattering medium. Afterwards, we formulate a statistical model that describes the probability density function of the light intensity, averaged over the ensemble of possible samples. Our predictions are validated in a set of experiments with first, second and third order feedback. 
\begin{figure}
		\centering
		\includegraphics[width=0.8\columnwidth]{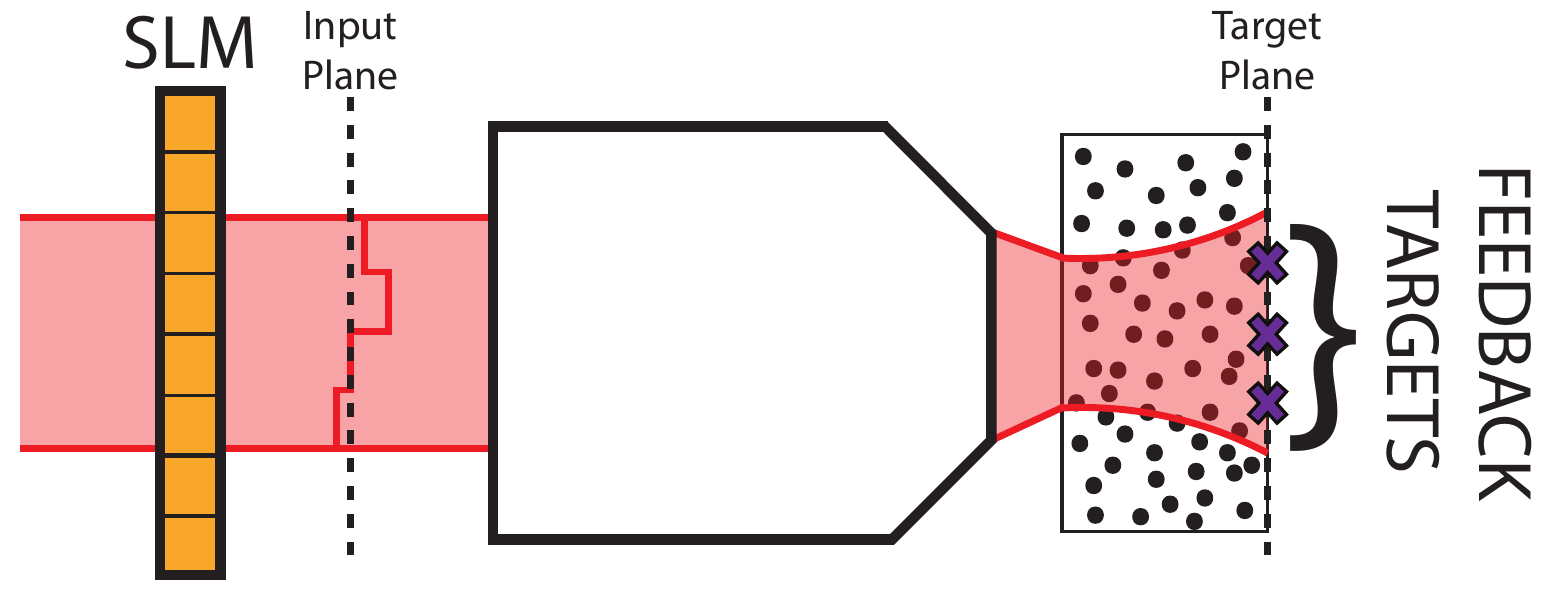}
		\caption{Illustration of the blind focusing experiment. SLM: Spatial light modulator.}
		\label{fig:blind_focusing}
\end{figure} 

\section{Theory of blind focusing}
Figure \ref{fig:blind_focusing} shows a simplified illustration of the blind focusing experiment. The goal of this experiment is to form a single diffraction limited focus through a scattering layer at the target plane using nonlinear feedback from $M$ targets. A spatial light modulator (SLM) is used to control the field $E_a$ at the input plane. The feedback signals from the individual targets cannot be distinguished and only the total generated signal is recorded. We define this feedback signal as
\begin{equation}
S \equiv \sum_b^M |E_b|^{2n},
\label{eq:feedback}
\end{equation}
where $n$ is the order of feedback, and $E_b$ represents the electric field at target $b$.

At the start of the experiment, an arbitrary starting wavefront is applied to the SLM. We then apply an optimization algorithm known as stepwise sequential wavefront shaping \cite{Vellekoop2015} to optimize the feedback signal. In contrast to the genetic algorithms used in other studies\cite{Katz2014,Conkey2012}, our method is deterministic and can be analyzed analytically.

Each iteration of the algorithm results in an optimized incident wavefront $E_a$, which is used as the starting wavefront of the next iteration of the algorithm. This process is repeated until the algorithm has reached a fixed point, where the optimized wavefront does not change anymore. In Appendix A, we prove that such a fixed point always corresponds to a (local) maximum of $S$.

In this section, we will derive how the target field $E_b$ changes as the $n$th-order feedback signal $S$ is optimized by iteratively running the algorithm. Segment by segment, this algorithm optimizes $S$ for a single incident field segments $E_a$, while keeping the other field segments static. By introducing the transmission matrix $\TM$, we can describe the target field as a linear combination of input fields
\begin{equation}
	E_b^{(k)} = \sum_a^N t_{ba} E_a^{(k)}.
    \label{eq:TM}
\end{equation}
Here, $t_{ba}$ is an element of the transmission matrix, $N$ is the number of independently controlled segments on the SLM and the superscript number $k$ between brackets indicates the iteration number. We take the total incident intensity $\sum_a^N |E_a|^2 = 1$.

For the $(k+1)$th wavefront shaping iteration, we use the optimized wavefront from the previous iteration as our new starting wavefront. Then we shift the phase of a single incident field segment a', such that $E^{(k)}_{a'}\rightarrow E^{(k)}_{a'}[\exp^{-i\phi} - 1]$.  We measure $S$ as we vary $\phi$ from $0$ to $2\pi$ in $P$ steps. Using Eq.~\eqref{eq:feedback} and Eq.~\eqref{eq:TM}, the intermediary feedback signal can now be written as
\begin{align}
		\tilde{S}^{(k+1)}(\phi) = \sum_b^M  \Big(|E_b^{(k)}|^2 + |t^{\supfill}_{ba'}E^{(k)}_{a'}|^2 |e^{i\phi}-1|^2  + ( E_b^{(k)}t^*_{ba'}E^{(k)*}_{a'} [e^{-i\phi}-1] + c.c.) \Big)^n,
        \label{eq:S_k+1}
\end{align}
where ${}^*$ represents the complex conjugate. Since we only change a single segment of the incident wavefront, the effect on $S$ is expected to be small. Therefore, we expand $\tilde{S}^{(k+1)}$ in terms of $E^{(k)}_{a'}[e^{-i\phi}-1]$ to arrive at
\begin{equation}
	\tilde{S}^{(k+1)}(\phi) \approx S^{(k)} + \sum_b^M  \Big( \weight^{(k)}_b t^*_{ba'} E_{a'}^{(k)*} [e^{-i\phi}-1] + c.c. \Big),
\label{eq:feedback2}
\end{equation}
with the nonlinearly weighted target field $\weight^{(k)}_b \equiv n |E_b^{(k)}|^{2(n-1)} E_{b}^{(k)}$ (see appendix A for a rigorous mathematical treatment).

Following the processing steps as described in \cite{Vellekoop2007}, we then find the optimized wavefront by isolating the contribution of $\phi$ using the following expression
\begin{equation}
	E^{(k+1)}_{a'} =  \frac{\norm^{(k+1)}}{{E}_{a'}^{(k)*} P}  \sum_{p}^{P} S(\phi_p) e^{i\phi_p} = \norm^{(k+1)} \sum_b^M  t^*_{ba'} \weight^{(k)}_b,
    \label{eq:optE_a1}
\end{equation}
where $\norm^{(k+1)}$ is a normalization factor, which normalizes total incident intensity again to 1. The phase shift $\phi$ at segment $a'$ is set back to 0 and the process is then repeated for all other SLM segments. Finally, by inserting Eq.~\eqref{eq:optE_a1} into Eq.~\eqref{eq:TM}, we arrive at an expression for the resulting target field after the $(k+1)$th wavefront shaping iteration
\begin{equation}
	E_{b}^{(k+1)} = \norm^{(k+1)} \sum_a^N \sum_{b'}^M t^{\supfill}_{ba} t^*_{b'a} \weight^{(k)}_{b'}.
    \label{eq:Ebopt}
\end{equation}
We recognize that the optimized target field in Eq.~\eqref{eq:Ebopt} is the nonlinearly weighted sum over the phase conjugated fields propagated from the target locations. The nonlinear weighted field scales with the field strength to the power $2n-1$, and thus the brighter targets will have a stronger contribution to the optimized incident field. Performing multiple iterations of wavefront can therefore ultimately result in a focus being formed at one of the targets. 

The expression in Eq.~\eqref{eq:Ebopt} can be connected to previous iterative wavefront shaping experiments. For linear feedback, $W_b^{(k)}$ reduces to $E_b^{(k)}$. In this case, after performing the wavefront shaping algorithm multiple times, $E_b^{(k)}$ will converge to the eigenvector of matrix $\TM$ with the highest eigenvalue \cite{Bosch2016}. In other words, instead of forming a focus, the algorithm will instead optimize the total transmission through the scattering sample. Alternatively, for $n>1$ in a weakly scattering medium, where $\TM$ is close to unitary, the optimized target field becomes $E_b^{(k+1)} \propto |E_b^{(k)}|^{2(n-1)}E^{(k)}_b$. Now the brightest targets are enhanced more than the dimmer targets until finally a focus is formed at the brightest target \cite{Papadopoulos2016}. However, we will show that when $\TM$ is not close to unitary, as is the case in strongly scattering media, maximizing $S$ is not necessary equivalent to forming a focus. 

\section{Statistical model for blind focusing through strongly scattering media}
Next, we want calculate the optimized target intensity distribution through a strongly scattering medium with a non-unitary transmission matrix. Finding the exact value of the target field, using Eq.~\eqref{eq:Ebopt}, requires full knowledge of the transmission matrix, which often cannot be obtained. Therefore, we will instead calculate the probability density function of $E_{b}^{(k+1)}$, averaged over the ensemble of possible samples. We assume that $W_{b}^{(k)}$ is known and independent of $\TM$. The transmission matrix is assumed to be a random matrix with independent elements, such that the expectation value $\langle t^{\supfill}_{ba} t^*_{b'a'} \rangle =\delta_{b'b} \delta_{a'a} \langle |t_{ba}|^2 \rangle$, where $\delta$ is the Kronecker delta. In Appendix B, we derive that for an imperfect wavefront shaping setup the probability density function of $E_b^{(k+1)}$ is a complex normal distribution of the form:
\begin{equation}
	P(E_{b}^{(k+1)}) = \frac{1}{\pi I_0} \exp \Bigg(- \frac{|E_{b}^{(k+1)}-\Emean_b^{(k+1)}|^2}{I_0} \Bigg).
    \label{eq:P(Eb)}
\end{equation}
Here, the average optimized field $\Emean_b^{(k+1)}$ and $I_0$ are given by
\begin{equation}
\Emean_b^{(k+1)} \equiv \frac{\weight_b^{(k)} }{\sqrt{\sum_{b'}^M |\weight_{b'}^{(k)}|^2}} \gamma \sqrt{N I_0} \qquad \text{and} \qquad I_0 \equiv \langle |t_{ba}|^2 \rangle,
\label{eq:Eb_mean}
\end{equation}
where the quality of the wavefront modulation is described by the fidelity parameter $|\gamma|^2$. This parameter's value ranges between 0 and 1, where a value of 1 corresponds with a perfect wavefront modulation. Unlike regular speckle fields, $E_{b}^{(k+1)}$ will have a non-zero average value $\Emean_b^{(k+1)}$ because of the performed wavefront shaping iteration. 

When $E_{b}^{(k+1)}$ follows a complex normal distribution with a non-zero mean, the corresponding optimized target intensity $I_b^{(k+1)} \equiv |E_{b}^{(k+1)}|^2$ follows a modified Rice distribution \cite{Goodman2007}. For $\Emean_b^{(k)} = 0$, this probability density function reduces to an exponential distribution as normally seen in speckle statistics. For now, we are mainly interested in the average optimized intensity and the corresponding standard deviation at target $b$, which are given by
\begin{align}\label{eq:Ibmean}
	\langle I_{b}^{(k+1)} \rangle = |\Emean_b^{(k+1)}|^2 + I_0 &= I_0 \Bigg(  N |\gamma|^2 \frac{|\weight_b^{(k)}|^2}{\sum_{b'}^M |\weight_{b'}^{(k)}|^2} + 1 \Bigg), \\
     \sigma_I =  \sqrt{I_0|\Emean_b^{(k+1)}|^2 + I^2_0} &= I_0 \sqrt{N |\gamma|^2 \frac{|\weight_b^{(k)}|^2}{\sum_{b'}^M |\weight_{b'}^{(k)}|^2} + 1},
    \label{eq:Ibstd}
\end{align}
respectively. Note that $\langle I_{b}^{(k+1)} \rangle$ increases with $N$, whereas $\sigma_I$ increases with $\sqrt{N}$. 

After the $(k+1)$th wavefront shaping iteration, the average intensity at target $b$  depends on the nonlinearly weighted fields at all targets. The targets which are generating a strong signal will contribute more to the feedback signal $S$, and will therefore, on average, be enhanced more than the weaker targets. We recognize that if nearly all of the feedback signal is coming from a single target, then the average intensity enhancement can be approximated by $\langle \eta \rangle \equiv \langle I_{b}^{(k+1)} \rangle / I_0 \approx N|\gamma|^2 + 1$. This expression is equivalent to the enhancement in a conventional single target wavefront shaping experiment \cite{Vellekoop2007}. However, when a large number of targets are contributing to the feedback signal, the expected enhancement is reduced to 1, and as a result, no focus is formed at all.

To summarize, in the blind focusing experiment, the statistics of the optimized intensities at the feedback targets are described by modified Rice distribution. Using Eq.~\eqref{eq:Ibmean}, we can predict how much the light intensity will be optimized at the feedback targets after an iteration of wavefront shaping.

\section{Experimental validation}
\subsection{Experimental setup}
\begin{figure}
		\centering
		\includegraphics[width=1\columnwidth]{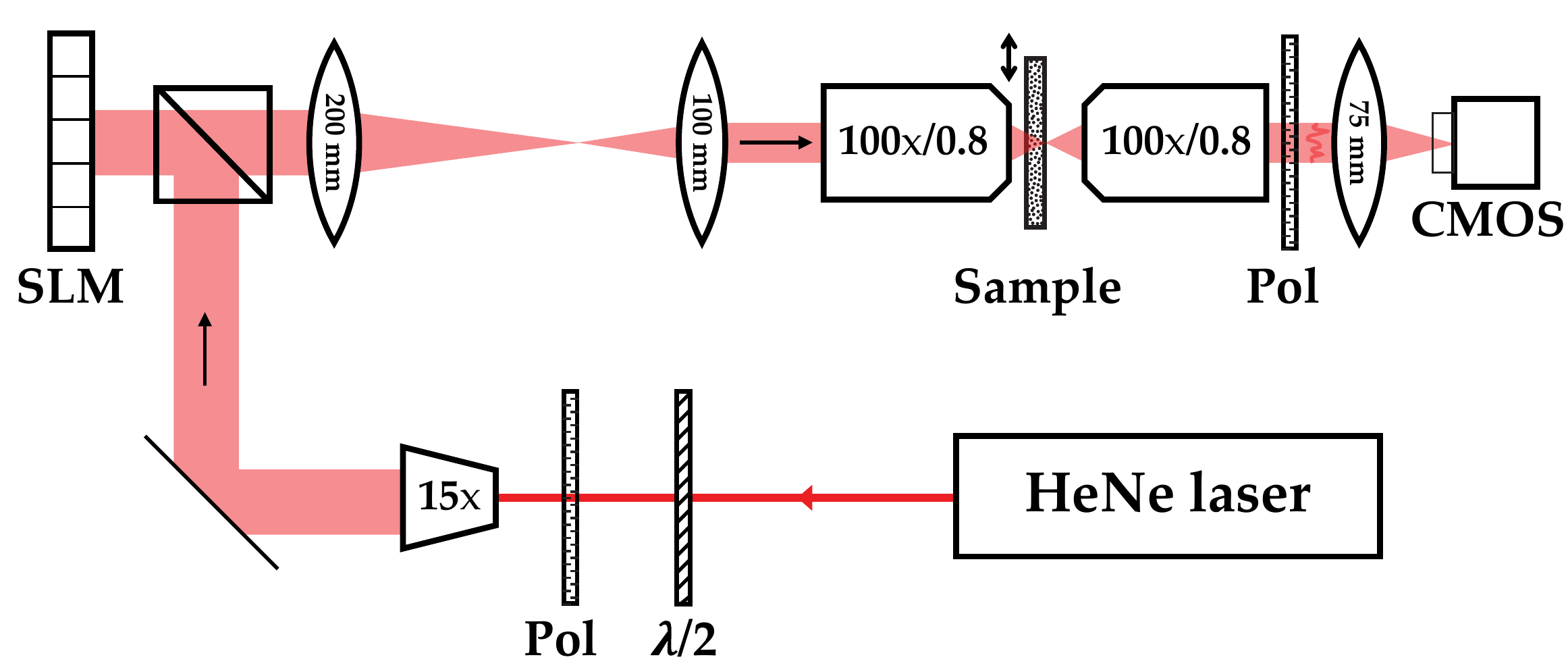}
		\caption{Schematic of the experimental setup. $\lambda$/2: half wave plate, Pol: Polarizer, SLM: Spatial light modulator, CMOS: Complementary metal oxide semiconductor camera.}
		\label{fig:experimental_setup}
\end{figure}

We will now experimentally validate the expected optimized intensity as predicted by Eq.~\eqref{eq:Ibmean}. Our experimental setup is illustrated in Fig.~\ref{fig:experimental_setup}. Light from a HeNe laser is expanded and modulated by a phase-only spatial light modulator (Hamamatsu X13138-07). With two lenses in a 4f-configuration, the SLM-modulated wavefront is conjugated to the back focal plane of a microscope objective (Zeiss A-Plan 100x/0.8), which focuses the light onto the surface of a scattering sample. As our source of feedback, we chose to use a CMOS camera (Basler acA640-750um), which measures the intensity distribution at the back surface of the scattering sample through an identical microscope objective. On the camera, we set groups of $3\times3$ pixels (corresponding to $0.16\times0.16$ $\mu$m$^2$) as independent targets for the wavefront shaping algorithm. Using this experimental setup, we can choose the order of feedback, and we can accurately set the number of targets contributing to the feedback signal. We would like to emphasize that in these experiments only the total $n$-th order intensities summed over all targets was used as the feedback signal, mimicking the nonlinear response of fluorescent markers in a multiphoton fluorescence excitation microscope. 

As a scattering sample, we used a glass substrate dip-coated in a suspension of 5$\%$ zinc-oxide (Sigma Aldrich, average grain size 200 nm) and demineralized water. The thickness of the layer were measured to be 37.6 $\pm$ 9.8 $\mu$m. Based on previous work \cite{Putten2011}, the transport mean free path of the zinc-oxide sample is expected to be around 0.6 $\mu$m, ensuring that all light passing through the sample is multiple scattered. The sample is mounted on a translation stage (Zaber T-LSM050A), to allow the interrogation of different sample locations for statistical averaging. The average wavefront shaping fidelity of our setup was measured to be $\langle |\gamma|^2 \rangle = 0.39$ with $N=208$.

\subsection{Blind focusing experiment with two feedback targets}
\begin{figure}
		\includegraphics[width=1\columnwidth]{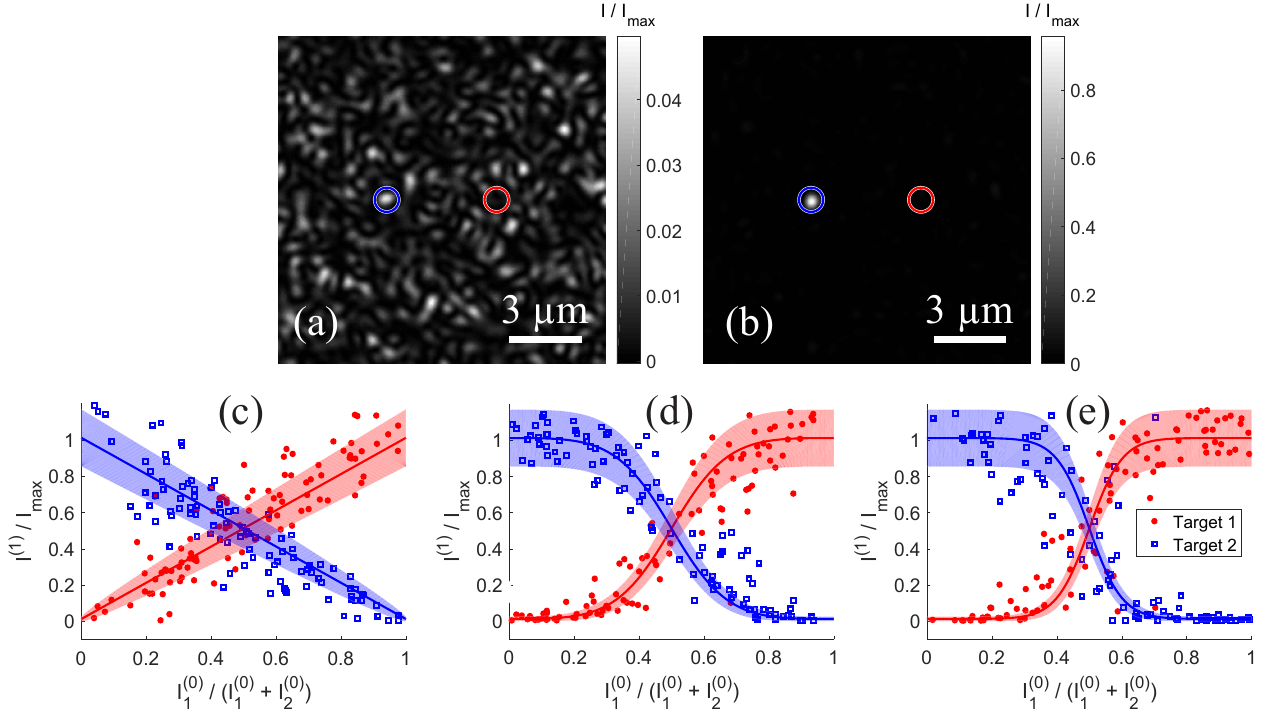}
		\caption{Results of the blind focusing experiment optimizing for two targets simultaneously. (a) Example image of the initial intensity distribution, and (b) intensity distribution after the first wavefront shaping iteration. The red (right) and blue (left) circles correspond to targets 1 and 2, respectively. The optimized intensities of target 1 (red circles) and target 2 (blue squares) are plotted as function of the initial ratio of the intensities in the two targets for (c) first-order, (d) second-order, and (e) third-order feedback. The predicted mean value for the optimized intensity and the corresponding standard deviation are represented by the colored solid lines and the shaded areas, respectively.}
		\label{fig:WFS_2targets}
\end{figure}

We performed a blind focusing experiment, where two targets were simultaneously optimized using the wavefront shaping algorithm as described before. The targets are separated by a distance 4.4 $\mu$m. On the SLM, a random pattern of $N=208$ square segments was displayed, matching the size of the light beam on the SLM. The experiment was performed 100 times, changing the initial SLM pattern and sample lateral position in between every experiment. In Fig.~\ref{fig:WFS_2targets}(a) and (b), examples of the intensity distributions measured on the back side of the sample, before (a) and after (b) the optimization, are shown. Here, targets 1 and 2 are indicated by the red (right) and blue (left) circle, respectively. In these figures, the starting intensity of target 2 is much higher than the intensity target 1. As a result, the contribution of target 2 to the feedback signal is much larger than target 1, and therefore, only the intensity of target 2 is enhanced. Fig.~\ref{fig:WFS_2targets}(c)-(e) show the optimized intensities $I^{(1)}$  of target 1 (red circles) and target 2 (blue squares) as function of the ratio of the initial intensities $I^{(0)}$ of the 2 targets, for first-order (c), second-order (d), and third-order (e) feedback. The optimized intensity is normalized by $I_{max} = N|\gamma|^2I_0$, which is the average intensity obtained by performing a single-target wavefront shaping experiment. The solid lines and the shaded areas indicate the average optimized intensities and the standard deviation as predicted by Eq.~\eqref{eq:Ibmean} and Eq.~\eqref{eq:Ibstd}.

In Fig.~\ref{fig:WFS_2targets}(c)-(e), we see that even when both targets contribute to the feedback signal, the target intensities are not necessarily equally enhanced. Neither are the optimized intensities $I^{(1)}$ randomly distributed, but rather the optimized intensities are directly related to the starting intensities of target 1 and 2, as expected from Eq.~\eqref{eq:Ibmean}. In the case of $n=1$, the optimized intensities at the targets are linearly proportional to the initial intensities, since in that case $|W_b^{(0)}|^2$ reduces to $I_b^{(0)}$. For higher-order feedback measurements, the relation between the initial intensity ratio and the optimized intensity becomes nonlinear. As a result, a small difference in $I^{(0)}$ between the two targets can result in a big difference between optimized intensities. For instance for $n = 3$, only approximately 65$\%$ of the total initial intensity is required in one target to ensure, in almost all cases, that only the intensity of that target will be optimized. This thresholding effect for $n>1$, observed in Fig.~\ref{fig:WFS_2targets}(d-e),  is the mechanism that allows the blind focusing method to form a single diffraction-limited focus even when multiple targets contribute to the feedback signal. The data is in good agreement with our theoretical predictions and most of the data points fall within the statistical variation as described in Eq.~\eqref{eq:Ibstd}. Other deviations might be explained by fluctuations in the fidelity during the experiment. 

\section{Blind focusing requirements}
\subsection{Theory}
Now, we want to use our statistical model to find out under which circumstances the blind focusing method is able to form a single focus, when $M$ targets are contributing to the feedback signal. Instead of analyzing convergence behaviour starting from a random speckle pattern, we analyze the case where light is already focused to one target before the optimization. The focus should be preserved after the wavefront shaping iteration when this focus corresponds to a fixed point of the algorithm. To quantify the intensity in the focus, we use the enhancement, which is defined as $\eta \equiv I/I_0$. The starting enhancement in the focus is given by $\eta^{(0)}$. Furthermore, we assume that all other $(M-1)$ targets are exponentially distributed with an average starting enhancement of $1$. Inserting these parameters into Eq.~\eqref{eq:Ibmean} produces an expression for the expected optimized focus enhancement after a single optimization iteration
\begin{equation}
	\langle \eta^{(1)} \rangle = N|\gamma|^2 \frac{(\eta^{(0)})^{2n-1}}{(\eta^{(0)})^{2n-1} + (2n-1)!(M-1)} + 1.
    \label{eq:eta_prediction}
\end{equation}

We recognize that $\langle \eta^{(1)} \rangle$ can be smaller than $\eta^{(0)}$ for a large $M$. In order for the blind focusing method to be able to form a focus, the average focus enhancement after the optimization should be equal to the enhancement before the optimization for some $\eta^{(0)} > 1$. In Appendix C, we show that (given $N$, $|\gamma|^2$ and $n$) an upper limit for the number of feedback targets can derived, which is given by
\begin{equation}
	M_{max} = \frac{(2n-2)^{2n-2}}{(2n-1)! (2n-1)^{2n-1} } (|\gamma|^2 N)^{2n-1} + 1.
    \label{eq:Mmax}
\end{equation}
Whenever $M > M_{max}$, a focus can on average not be formed. Rather, the wavefront shaping algorithm optimizes the intensity and contrast of the full speckle pattern to maximize the feedback signal. We see that the order of feedback $n$ can be increased to guarantee blind focusing convergence for a larger number of contributing targets. Moreover, by increasing $N$ by a factor of $\alpha$, $M_{max}$ increases by a factor of $\alpha^{2n-1}$. 

\subsection{Blind focusing experiment with $M$ targets}
To verify the prediction in Eq.~\eqref{eq:eta_prediction}, we performed a second experiment using a large feedback area, containing $M$ feedback targets. For this experiment, the same experimental setup was used as in the first experiment. We start our experiment with a pre-optimized focus at a single target, whereas the remaining targets in the feedback area are illuminated by a random speckle pattern. The pre-optimized focus is obtained by first optimizing for a single target in the center of the camera frame. Afterwards, a wavefront shaping iteration is performed using the total second-order feedback signal from all targets within a small and a large region of interest (ROI), which have a radius of 2.1 $\mu$m and 10.5 $\mu$m. Based on the average speckle size, the number of targets in the ROIs are estimated to be $M=96$ and $M=2400$ for the small and the large ROIs, respectively. The blind focusing experiments were performed for $N=80$ and $N=208$, and were performed 100 times for both ROIs. In between each experiment, the lateral sample position was changed and the intensity of pre-optimized focus was varied by adding a controlled amount of uniformly-distributed noise to the starting wavefront. 

\begin{figure}
		\includegraphics[width=1\textwidth]{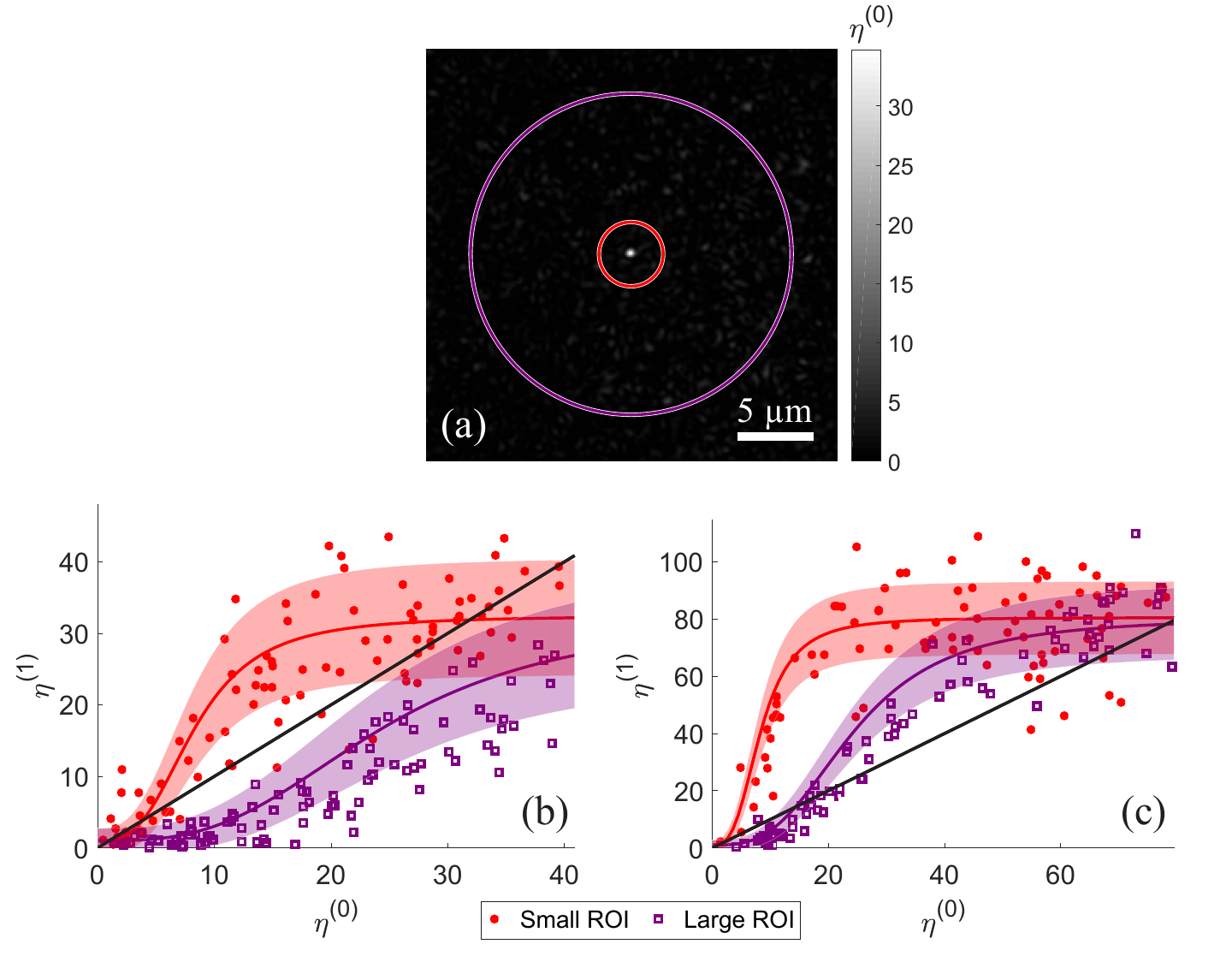}
		\caption{Results of the blind focusing experiments with a pre-optimized focus using the total second-order feedback signal from $M$ targets. (a) Example speckle pattern with a pre-optimized focus, where the red and purple circles represent the small ($M=96$) and large ($M=2400$) regions of interest, respectively. (b-c) The enhancement $\eta$ in the pre-optimized focus before and after blind focusing with $M=96$ (red circles) and $M=2400$ (purple squares) with (b) $N = 80$ and (c) $N=208$. The predicted mean value for $\eta^{(1)}$ and the corresponding standard deviation are represented by the colored solid lines and the shaded areas, respectively. The black solid line indicates the identity line.}
		\label{fig:bf_preoptimized}
\end{figure}

In Fig.~\ref{fig:bf_preoptimized}(a), an example image of a starting speckle pattern with a pre-optimized focus is shown. The red and purple circles in the figure indicate the size of the small and large ROIs. In Fig.~\ref{fig:bf_preoptimized}(b) and (c), the enhancement of the pre-optimized focus intensity after the blind focusing experiment $\eta^{(1)}$ is plotted as function of the starting enhancement of the pre-optimized focus, for $N=80$ (Fig.~\ref{fig:bf_preoptimized}b) and $N=208$ (Fig.~\ref{fig:bf_preoptimized}c). The experiments with small and the large ROI are represented by the red circles and the purple squares, respectively. The expected value for $\eta^{(1)}$, as predicted by Eq.~\eqref{eq:eta_prediction}, and the corresponding standard deviation are represented in Fig.~\ref{fig:bf_preoptimized}(b) and (c) by the colored solid lines and the shaded areas, respectively. The black solid lines indicates the identity lines, where $\eta^{(1)} = \eta^{(0)}$. 

In Fig.~\ref{fig:bf_preoptimized}(b) and (c), two general regions can be recognized. The data points above the identity line represent experiments where the focus enhancement increased, whereas the data points under the identity line represent a decrease in the focus enhancement. In Fig.~\ref{fig:bf_preoptimized}(b), in the experiments with $M = 96$ and $N=80$, nearly all measured $\eta^{(1)}$ values lie above the identity line, meaning that the focus enhancement increases after the iteration of wavefront shaping. However, in the experiments with the large ROI, $\eta^{(1)}$ is almost always lower than $\eta^{(0)}$. These results suggest that the large ROI contains too many competing targets for the pre-optimized focus to be preserved. In other words, the wavefront shaping algorithm favors optimizing the intensity of multiple targets within the ROI over preserving the intensity in the pre-optimized focus. 

In Fig.~\ref{fig:bf_preoptimized}(c), the number of controlled SLM segments was increased to $N =208$. In this figure, the $\langle \eta^{(1)} \rangle$ curves cross the identity curves for both the small and large ROIs, indicating that the intensity in the pre-optimized focus was further enhanced. Increasing $N$ thus allows blind focusing to preserve a focus for a higher number of targets that are contributing to the feedback signal. The experimental data is in excellent agreement with the prediction in Eq.~\eqref{eq:eta_prediction}.

These results demonstrate that for second-order feedback, the blind focusing method will not form a focus using only 80 SLM segments when an area with 2400 targets is illuminated. This is in accordance with the prediction in Eq.~\eqref{eq:Mmax}, since the number of feedback targets in the large ROI ($M=2400$) exceeds the upper limit, $M_{max} =  751$, whereas the experiment with $N = 208$ (in Fig.~\ref{fig:bf_preoptimized}(c)) has a higher $M_{max}$ of 13181 targets. 



\section{Discussion}
We studied wavefront shaping with nonlinear feedback through a strongly scattering sample using the stepwise sequential algorithm \cite{Vellekoop2007}. This deterministic algorithm allowed us to derive the exact solution for the optimized field at the feedback targets, which is given in Eq.~\eqref{eq:Ebopt}. This expression shows that in weakly scattering samples, the optimization algorithm takes higher orders of the initial target field with every iteration. As a result, only the brightest speckles at the target plane will be enhanced \cite{Papadopoulos2016}. Therefore, wavefront shaping and adaptive optics techniques can be used to improve the intensity of the focus in multiphoton fluorescence excitation microscopy without the need for a guide star \cite{Tang2012,Sinefeld2015}. 

Katz et al. \cite{Katz2014} have shown that a focus can be formed through a strongly scattering layer without the need of a localized reporter. We showed that in strongly scattering samples, the statistics of the optimized target intensities are accurately described by the Rice distribution. However, due to the large standard deviation in the distribution, it remains hard to predict at which target the focus will be formed when the targets are illuminated with a random speckle pattern.

In Appendix A, we showed that a fixed point of our optimization algorithm always corresponds to a maximum of the feedback signal. Moreover, we experimentally demonstrated that when the number of targets exceeds $M_{max}$ from Eq.~\eqref{eq:Mmax}, our algorithm is, on average, unable to form a focus. As a result, maximizing $S$ does not always guarantee the formation of a focus, but will instead optimize the intensity and contrast of the speckle pattern. This limitation applies to all optimization algorithms using the total nonlinear signal as feedback, including the commonly used genetic wavefront shaping algorithm \cite{Conkey2012}.

In our experiments, we only considered targets distributed across a flat two-dimensional image plane. However, our model can also be employed to point targets on an arbitrarily shaped plane. Therefore, our model can easily be generalized for structure with targets distributed in three spatial dimensions. In such samples, the light intensity will spread out more the deeper the light propagates into the sample.  As a result, the number of illuminated targets will also rapidly increase with imaging depth. 

In our statistical model, we assumed that the nonlinearly weighted field $W_b^{(k)}$ is independent of the transmission matrix. As a result, we ignore certain properties of the matrix $\TM$ known from random matrix theory, such as the distribution of transmission eigenvalues \cite{Rotter2017}. Nevertheless, our statistical model is in a surprisingly good agreement with the experimental data. At the moment, our model is unable to predict the rate of convergence. Therefore, we believe that studying these properties of random matrices will be interesting to get a full understanding of this nonlinear optimization process.


\section*{Appendix A: Convergence to local maximum}
We used the stepwise sequential optimization algorithm \cite{Vellekoop2015} in an attempt to maximize the total feedback signal $S$. For linear feedback from a single target, it is well known that this algorithm finds a global maximum for the intensity at that target in a single iteration \cite{Vellekoop2007}. In the case of nonlinear feedback originating from multiple targets, however, the optimization problem has multiple local maxima. In this situation, it is not directly trivial that the algorithm finds a local maximum of $S$; it is not even directly clear that the stepwise sequential algorithm increases $S$ at all.

Here, we will prove that all local maxima of $S$ correspond to attractive fixed points of the algorithm (Eq.~\eqref{eq:Ebopt}) and vice versa.
Consequentially, when the algorithm converges, it converges to a local maximum of $S$ as well. Furthermore, this proof implies that each local maximum of $S$ has a finite region of attraction for which the algorithm will converge to that local maximum.
Note that we cannot exclude the existence of initial conditions for which the algorithm does not converge. However, we have not observed these cases in our experiments.
In order to simplify the derivations, we introduce a compact vector notation, replacing $E_a$ by $\aa$, $E_b$ by $\vv{b}$ etc. For example, Eq.~\eqref{eq:TM} can now be written
\begin{equation}\tag{A1}
    \vv{b} = \TM \aa,
\end{equation}
with $\TM$ the transmission matrix. The incident field is normalized so that $\pnorm{\aa}=1$.

\subsection*{Wirtinger calculus}
To analyze convergence to a local maximum, we will expand the system around a fixed point $\vv[*]{a}$ . Here a technical complication arises: due to the complex conjugate in 
$S$, it is not a holomorphic function of $\aa$, so the complex derivative $\partial{S}/\partial{\aa}$ does not exist.

In order to avoid this complication, we use Wirtinger calculus to calculate the derivatives \cite{Fisher2002}. In Wirtinger calculus, $S(\aa)$ is replaced by a function $S(\aa,\ac)$, where $\aa$ and $\ac$ are \emph{independent} variables: technically, $\ac$ is \emph{not} equal to the complex conjugate of $\aa$. However, as long as we restrict ourselves to only evaluate $S(\aa,\ac)$ in the subspace $\ac=\aa^*$, we can perform all derivatives in the usual manner and obtain correct results.
Using the compact notation and Wirtinger calculus, we can now write Eq.~\eqref{eq:feedback} as 
\begin{equation}\tag{A2}
    S(\aa, \ac) = \sum_b^M \left[(\TM^* \ac)_b (\TM \aa)_b\right]^n,
\end{equation}
where $()_b$ denotes the $b$-th element of the vector in the parentheses. As an example, we use Wirtinger calculus to calculate the first order Taylor expansion of $S$ for small perturbations $\vv{\Delta}$ around $\aa$
\begin{equation}\tag{A3}
    S(\aa+\DD, \ac+\Dc) = S(\aa,\ac) + \pderiv{S}{\aa}\DD + \pderiv{S}{\ac}\Dc + O(\pnorm{\DD}^2).
    \label{eq:S-a-Taylor}
\end{equation}
We can now evaluate the derivative towards component $a_i$ of the incident field $\aa$
\begin{equation}\tag{A4}
    \pderiv{S}{a_i} = n\sum_b^M \left[(\TM^* \ac)_b (\TM \aa)_b\right]^{n-1} t_{bi} = \sum_b^M w_b^* t_{bi},
\end{equation}
where the elements of $\ww$ and $\wc$ are given by
\begin{equation}\tag{A5}
    w_b(\aa,\ac) = n(\TM \aa)_b^n (\TM^* \ac)_b^{(n-1)} \qquad\text{and}\qquad
    \overline{w}_b(\aa,\ac) = n(\TM^* \ac)_b^n (\TM \aa)_b^{(n-1)}.
    \label{eq:w-elem}
\end{equation}
By evaluating the derivative for all elements $a_i$, we get the vector derivatives
\begin{align}\tag{A6}
    \pderiv{S}{\aa} = \wwd \TM\qquad\text{and}\qquad
    \pderiv{S}{\ac} = \ww^T \TM^*.
    \label{eq:dS-da}
\end{align}
We can use the Taylor expansion Eq.~\eqref{eq:S-a-Taylor} to calculate what happens during the optimization process, when the phase of a single element of the incident field is changed. The perturbation corresponding to changing the phase of segment $a'$ is given by 
\begin{equation}\tag{A7}
    \Delta_i =
    \begin{cases}
    a_i (e^{-i\phi}-1)&\text{for }i=a'\\
    0&\text{otherwise}
    \end{cases}.
\end{equation}
Inserting $\DD$ into Eq.~\eqref{eq:S-a-Taylor} gives Eq.~\eqref{eq:feedback2} from the main text.

\subsection*{Attractivity of fixed point}
The algorithm in Eq.~\eqref{eq:optE_a1} can be thought of as a cyclical series of mappings $\aa[(k)] \rightarrow \vv[(k)]{b} \rightarrow \vv[(k)]{w} \rightarrow \TMd \vv[(k)]{w} \rightarrow \vv[(k+1)]{a}$, etc., where $\TMd$ is the conjugate transpose of $\TM$. We rewrite Eq.~\eqref{eq:optE_a1} as
\begin{equation}\tag{A8}
\ff(\aa, \ac) = \frac{\TMd\vv{w}(\aa, \ac)}{\pnorm{\TMd\vv{w}(\aa, \ac)}},
\label{eq:f-definition}
\end{equation}
where $\ff$ now defines the mapping from $\aa[(k)]$ to $\aa[(k+1)]$. Using Wirtinger calculus, we can make a first order approximation of the mapping for a small perturbation $\vv{\Delta}$ around a fixed point of the mapping, $\af=\ff(\af)$. 
\begin{equation}\tag{A9}
\ff(\af+\DD, \afc+\Dc)
\approx
\ff(\af,\afc)
+
\begin{bmatrix}
\pderiv{\ff}{\aa} & \pderiv{\ff}{\ac}\\
\pderiv{\fc}{\aa} & \pderiv{\fc}{\ac}
\end{bmatrix}_\fsuba
\begin{bmatrix}
\DD \\ \Dc
\end{bmatrix},\label{eq:Jacobian}
\end{equation}
where the matrix is $J_f(\af)$: the Jacobian of $\ff$, evaluated at the fixed point $\af$. When the fixed point is attractive in some finite region, each iteration brings $\aa$ closer to $\af$, so we must have $\pnorm{\aa[(k+1)]-\af} \leq q\pnorm{\aa[(k)]-\af}$, with $0\leq q<1$. From Eq.~\eqref{eq:Jacobian}, we see that this condition is equivalent to saying that the spectral radius of the Jacobian $\rho(J_f)<1$. We will show below that this condition is always met at local maxima of $S$.

From the definition of $\ff$ it is clear that any perturbation in the direction of $\af$ will have no effect at all. Therefore, we restrict ourselves to perturbations perpendicular to $\af$, i.e. $\Dc^T \af=0$, hence $\Dc^T \TMd \wf=0$. Under this condition, we can find a simple expression for terms of the form $\Dc^T \partial{\ff}/\partial{\aa}$
\begin{equation}\tag{A10}
    \Dc^T\valueat{\pderiv{\ff}{\aa}}{\fsuba} =\Dc^T\left[\frac{\TMd}{\pnorm{\TMd\vv{w}}}\pderiv{\vv{w}}{\aa} + \TMd\vv{w}\pderiv{}{\aa}\frac{1}{\pnorm{\TMd\vv{w}}}\right]_\fsuba
    =\Dc^T \frac{\TMd}{\pnorm{\TMd\vv{w}}}\valueat{\pderiv{\ww}{\aa}}{\fsuba},
    \label{eq:D-df-da}
\end{equation}
which we will use in the next section. Here the product rule was used in the first step, and orthogonality of $\DD$ and $\af$ was used in the second step.

\subsection*{Local maximum}
To find local maxima of $S$ under the constraint that $\pnorm{\aa}=1$, we apply the method of Lagrange multipliers and minimize the Lagrange function
\begin{equation}\tag{A11}
    S_L(\aa, \ac) = S(\aa, \ac) - \lambda (\ad \aa - 1),
\end{equation}
where $\lambda$ is a Lagrange multiplier and $\ad \aa - 1$ represents the constraint that $\pnorm{\aa}=1$.

The first order conditions for a local maximum follow by equating the first derivatives of $S_L$ to zero, giving
\begin{align}\tag{A12}
    \wfd \TM - \lambda \afd &= 0\\\tag{A13}
    \wf^T \TM^* - \lambda \af^T &= 0\\\tag{A14}
    \afd \af -1 &= 0,
\end{align}
with the solution $\lambda=\pnorm{\TMd \wf}$ and $\af = \TMd \wf / \lambda$, proving that a fixed point of $\vv{f}$ is also a critical point of $S_L$. In order to prove that this stationary point is a local maximum (and not a local minimum or a saddle point), we need show that any small perturbation that maintains the first order conditions decreases the value of $S_L$. In order to do so, we evaluate the second derivative (the Hessian matrix $\vv{H}$) of $S_L$ \emph{considering only perturbations perpendicular to $\af$}, i.e. perturbations that maintain the constraint $\pnorm{\aa}=1$ to the first order. Using Eq.~\eqref{eq:dS-da}, we arrive at
\begin{equation}\tag{A15}
    \vv{H} \equiv
    \begin{bmatrix}
    \ppderiv{S_L}{\aa}{\ac} & \ppderivs{S_L}{\ac}\\
    \ppderivs{S_L}{\aa} & \ppderiv{S_L}{\ac}{\aa}
    \end{bmatrix}_\fsuba
    =
    \begin{bmatrix}
    \TMd \pderiv{\ww}{\aa} - \lambda& \TMd \pderiv{\ww}{\ac}\\
    \TM^T \pderiv{\wc}{\aa} & \TM^T \pderiv{\wc}{\ac} - \lambda
    \end{bmatrix}_\fsuba.
\end{equation}
Using Eq.~\eqref{eq:D-df-da}, we find
\begin{equation}\tag{A16}
    \begin{bmatrix}\Dc \\ \DD\end{bmatrix}^T
    \vv{H}
    \begin{bmatrix}\DD \\ \Dc\end{bmatrix} =
    \begin{bmatrix}\Dc \\ \DD\end{bmatrix}^T
    \lambda(J_f - \vv{I})
    \begin{bmatrix}\DD \\ \Dc\end{bmatrix}<0,
    \label{eq:H-negative-proof}
\end{equation}
with $\vv{I}$ the identity matrix. Since $\rho(J_f)<1$, from Eq.~\eqref{eq:H-negative-proof}, $J_f-\vv{I}$ is negative definite, proving the last step: every perturbation $\Delta$ decreases the signal. Hence, every fixed point $\af$ corresponds to a local maximum of the constrained optimization problem.

\subsection*{Conclusion}
This final result in Eq.~\eqref{eq:H-negative-proof} shows that any small perturbation around the fixed point decreases the signal. In conclusion, we demonstrated that, when $\af$ is an attractive fixed point of mapping $\ff$, the resulting signal $S(\af, \afc)$ \emph{must} be a local maximum. The converse is also true, since both statements are equivalent to the condition that $\rho(J_f)<1$. Finally, as detailed in the main text, even though the algorithm maximizes $S$, such a maximum does not necessarily correspond to a focus.

\section*{Appendix B: Derivation of the blind focusing statistical model}

In this appendix, we derive Eq.~\eqref{eq:P(Eb)}, the complex normal distribution of the optimized target field when blind focusing through a strongly scattering sample. We start by constructing the optimized incident field, which is given in Eq.~\eqref{eq:optE_a1}. In an imperfect wavefront shaping setup, the quality of the wavefront modulation is described by the fidelity, $|\gamma|^2$. The constructed input field is given by $\hat{E}^{(k+1)}_a = \gamma E^{(k+1)}_a + \sqrt{1-|\gamma|^2} \Delta E_a$, where $E^{(k+1)}_a$ is the desired input field and $\Delta E_a$ is a normalized field, which is by definition orthogonal to $E_a^{(k+1)}$ \cite{Vellekoop2008}. When this imperfect input field is inserted into Eq.~\eqref{eq:TM}, we find that the constructed optimized target field becomes
\begin{equation}\tag{B1}
	\hat{E}_{b}^{(k+1)} = \gamma  \norm^{(k+1)} \sum_a^N \sum_{b'}^M t^{\supfill}_{ba} t^*_{b'a} \weight^{(k)}_{b'} +  \sqrt{1-|\gamma|^2} \zeta_b.
    \label{eqB:Ebopt}
\end{equation}
Here, $\zeta_b \equiv \sum_a^N t_{ba} \Delta E_a$ which is assumed to be an uncorrelated scattered field with $\langle \zeta_b \rangle = 0$ and $\text{var}(\zeta_b) = \langle|t_{ba}|^2\rangle$. Note that for a perfect setup with $\gamma = 1$, $\hat{E}_{b}^{(k+1)} = E_{b}^{(k+1)}$ (Eq.~\eqref{eq:Ebopt}). We recognize that Eq.~\eqref{eqB:Ebopt} can be written as a sum over $N$ independent random variables $\chi_a$
\begin{equation}\tag{B2}
	\hat{E}_{b}^{(k+1)} = \gamma \norm^{(k+1)} \sum_a^N \chi^{(k)}_a + \sqrt{1-|\gamma|^2} \zeta_b \qquad\textrm{with}\qquad\chi^{(k)}_a\equiv\sum_{b'}^M t^{\supfill}_{ba} t^*_{b'a} \weight^{(k)}_{b'}.
\label{eqB:Eb_chi}
\end{equation}
When $N$ is large, by the central limit theorem $\hat{E}_b^{(k+1)}$ has a complex normal distribution.

To find the average and variance of $\hat{E}_b^{(k+1)}$, we start by calculating the first and second raw moments of the terms $\chi_a$. As stated in the main text, we assume that $W_{b}^{(k)}$ is known and independent of $\TM$, and that $\langle t^{\supfill}_{ba} t^*_{b'a'} \rangle =\delta_{b'b} \delta_{a'a} \langle |t_{ba}|^2 \rangle$. Under these assumptions, the first moment is given by 
\begin{equation}\tag{B3}
\langle\chi^{(k)}_a\rangle = \sum_{b'}^M \weight^{(k)}_{b'} \langle t^{\supfill}_{ba} t^*_{b'a} \rangle
= \weight^{(k)}_{b} \langle|t_{ba}|^2\rangle,
\label{eqB:chi_mean}
\end{equation}
and the second moment
\begin{align}\tag{B4}
\langle|\chi^{(k)}_a|^2\rangle &= \sum_{b'}^M \sum_{b''}^M \weight^{(k)}_{b'} \weight^{(k)*}_{b''} \langle |t^{\supfill}_{ba}|^2 t^*_{b'a} t^{\supfill}_{b''a}\rangle\\ \tag{B5}
&= \sum_{b'}^M |\weight^{(k)}_{b'}|^2 \langle |t_{ba}|^2 |t_{b'a}|^2 \rangle\\ \tag{B6}
&= \sum_{b'\neq b}^M |\weight^{(k)}_{b'}|^2 \langle|t_{ba}|^2\rangle \langle|t_{b'a}|^2\rangle + |\weight^{(k)}_{b}|^2\langle|t_{ba}|^4\rangle.
\end{align}
Realizing that for a Gaussian distribution, $\langle|t_{ba}|^4\rangle=2 \langle|t_{ba}|^2\rangle^2$, we find
\begin{equation}\tag{B7}
\text{var}(\chi^{(k)}_a) =  \langle|\chi^{(k)}_a|^2\rangle-|\langle\chi^{(k)}_a\rangle|^2 = \langle|t_{ba}|^2\rangle^2 \sum_{b'}^M |\weight^{(k)}_{b'}|^2.
\label{eqB:chi_var}
\end{equation}

Next, we calculate the value of $\norm^{(k+1)}$, which was introduced in Eq.~\eqref{eq:optE_a1} to normalize the total incident intensity after the optimization. We assume that, for large $N$, the normalization factor is self-averaging, such that
\begin{align}\tag{B8}
\norm^{(k+1)} &=  \sqrt{\frac{1}{\sum_a^N |\sum_{b}^M t^*_{ba}\weight^{(k)}_{b}|^2}} \approx \sqrt{\frac{1}{N\langle|t_{ba}|^2\rangle \sum_{b}^M |\weight^{(k)}_{b}|^2}}.
\end{align}

To obtain the mean of $\hat{E}_b^{(k+1)}$, we can simply add the means of $\chi_a$ and $\zeta_b$, since the two variables are uncorrelated. We calculate the optimized field average by inserting Eq.~\eqref{eqB:chi_mean} and $\langle\zeta^{(k)}_b\rangle = 0$ into Eq.~\eqref{eqB:Eb_chi}
\begin{equation}\tag{B9}
\langle \hat{E}_b^{(k+1)} \rangle = \gamma \norm^{(k+1)} \sum_a^N \langle\chi^{(k)}_a\rangle + \sqrt{1-|\gamma|^2} \langle \zeta_b \rangle = \frac{W_b^{(k)}}{\sqrt{\sum_{b'}^M |W_{b'}^{(k)}|^2}} \gamma \sqrt{N\langle|t_{ba}|^2\rangle }.
\end{equation}
Similarly, we can calculate the variance of $\hat{E}_b^{(k+1)}$ by adding $\text{var}(\chi^{(k)}_a)$ (Eq.~\eqref{eqB:chi_var}) and $\text{var}(\zeta_b) = \langle|t_{ba}|^2\rangle$ in the following manner
\begin{align} \tag{B10}
\text{var} (\hat{E}_b^{(k+1)}) &= |\gamma|^2  ( \norm^{(k+1)} )^2 \sum_a^N \text{var}(\chi^{(k)}_a) + \Big(1-|\gamma|^2\Big) \text{var} (\zeta_b) \\ \tag{B11}
&= |\gamma|^2 \frac{N \langle|t_{ba}|^2\rangle^2 \sum_{b'}^M |\weight^{(k)}_{b'}|^2}{N\langle|t_{ba}|^2\rangle \sum_{b''}^M |\weight^{(k)}_{b''}|^2} + \Big(1-|\gamma|^2\Big) \langle|t_{ba}|^2\rangle \\ \tag{B12}
&= \langle|t_{ba}|^2\rangle.
\end{align}
When we substitute $\mu_b^{(k+1)} \equiv \langle \hat{E}_b^{(k+1)} \rangle$ and $I_0 \equiv \text{var} (\hat{E}_b^{(k+1)})$, we arrive at the expressions found in Eq.~\eqref{eq:Eb_mean}.

\section*{Appendix C: Derivation of the blind focusing requirements}
In this section, we calculate the minimal requirements for the formation of a focus using the blind focus method. We assume that, when a focus is formed, only the intensity at the focus location is enhanced and that all other targets have an exponentially distributed enhancement with an average of 1. A focus can be formed when the expected focus enhancement of the next iteration $\langle \eta^{(1)} \rangle$ (as described by Eq.~\eqref{eq:eta_prediction}) is larger than or equal to the current focus enhancement, for some $\eta^{(0)} > 1$. For simplification, we instead consider $\langle \eta^{(1)} \rangle \geq \eta^{(0)} + 1$, which is a slightly more restrictive condition. We can write this condition as
\begin{align}\tag{C1}
    N|\gamma|^2 \frac{(\eta^{(0)})^{2n-1}}{(\eta^{(0)})^{2n-1} + (2n-1)!(M-1)} + 1 &\geq \eta^{(0)} + 1 \\  \tag{C2}
    -(\eta^{(0)})^{2n-1} + N|\gamma|^2 (\eta^{(0)})^{2n-2} &\geq (2n-1)! (M-1).
    \label{eqC:condition}
\end{align}

Next, we derive the minimum requirements for this condition to be satisfied. Therefore, we proceed by finding
\begin{equation}\tag{C3}
    \max_{\eta^{(0)}} \Big(-(\eta^{(0)})^{2n-1} + N|\gamma|^2 (\eta^{(0)})^{2n-2} \Big),
    \label{eqC:max}
\end{equation}
to find out in which cases this maximum value satisfies the condition in Eq.~\eqref{eqC:condition}. We maximize this function by equating the derivative of this function towards $\eta^{(0)}$ to zero, giving
\begin{align}\tag{C4}
    (2n-1)(\eta^{(0)})^{2n-2} &= (2n-2)N|\gamma|^2 (\eta^{(0)})^{2n-3} \\ \tag{C5}
    \eta^{(0)} &=  \frac{2n-2}{2n-1} N|\gamma|^2.
\end{align}
We insert this maximized value for $\eta^{(0)}$ into Eq.~\eqref{eqC:condition}
\begin{equation}\tag{C6}
    -\Big({\frac{2n-2}{2n-1} N|\gamma|^2}\Big)^{2n-1} + N|\gamma|^2 \Big({\frac{2n-2}{2n-1} N|\gamma|^2}\Big)^{2n-2} \geq (2n-1)! (M-1).
    \label{eqC:polynomial2}
\end{equation}
Given $N$, $|\gamma|^2$ and $n$, we solve Eq.~\eqref{eqC:polynomial2} for $M$
\begin{align} \tag{C7}
      (2n-1)! (M-1) &<  \Big(\frac{2n-2}{2n-1}\Big)^{2n-2} \Big(1 - \frac{2n-2}{2n-1}\Big) (N|\gamma|^2)^{2n-1} \\ \tag{C8}
    (2n-1)! (M-1) &<   \frac{(2n-2)^{2n-2}}{(2n-1)^{2n-1}} (N|\gamma|^2)^{2n-1} \\ \tag{C9}
    M &<  \frac{(2n-2)^{2n-2}}{(2n-1)! (2n-1)^{2n-1} } (|\gamma|^2 N)^{2n-1} + 1.
\end{align}
This final expression is the upper-limit for $M$, which corresponds to the parameter $M_{max}$ as given in  Eq.~\eqref{eq:Mmax}. Whenever $M<M_{max}$, the condition in Eq.~\eqref{eqC:condition} will be satisfied, which means that, on average, blind focusing is indeed able to form a a single focus.

\section*{Funding}
This work was funded by the European Research Council under the European Union's Horizon 2020 (ERC-2016-StG-678919).

\section*{Acknowledgements}
The authors would like to thank: Yoeri Boink for the helpful discussions and Tom Knop for his aid in the fabrication of the samples.


\end{document}